\documentclass[preprint,amsmath,showpacs]{revtex4}
\begin{document}

\title{Absence of localization in a disordered one-dimensional ring threaded by an Aharonov-Bohm flux}
\author{Jean Heinrichs}
\email{J.Heinrichs@ulg.ac.be} \affiliation{D\'{e}partement de
Physique, B{5a}, Universit\'{e} de Li\`{e}ge, Sart Tilman, B-{4000}
Li\`{e}ge, Belgium}
\date{\today}

\begin{abstract}
Absence of localization is demonstrated analytically to leading
order in weak disorder in a one-dimensional Anderson model of a ring
threaded by an Aharonov-Bohm (A-B) flux.  The result follows from
adapting an earlier perturbation treatment of disorder in a
superconducting ring subjected to an imaginary vector potential
proportional to a depinning field for flux lines bound to random
columnar defects parallel to the axis of the ring.  The absence of
localization in the ring threaded by an A-B flux for sufficiently
weak disorder is compatible with large free electron type persistent
current obtained in recent studies of the above model.
\end{abstract}


\maketitle

The nature (localized or delocalized) of the eigentstates in a
disordered metallic ring threaded by an Aharonov-Bohm (A-B) flux is
expected to strongly influence the magnitude of the persistent
current\cite{1} which may be induced in the ring.  In recent years
the existence of persistent current has been demonstrated
experimentally in metallic and semiconducting mesoscopic ring
systems\cite{2,3,4,5,6}.The various theoretical predictions of the
magnitude of persistent current are, however, stagnating around
values lying between one- and two orders of magnitude below the
experimental results\cite{7}.

While in a disordered linear chain (and, likewise, in an infinite
disordered ring as well) all eigenstates are localized\cite{8,9},
the precise effect of an A-B flux on these states in a ring has, to
out knowledge, not been discussed in detail so far.  The aim of this
note is to analyse the problem, using a simple extension of earlier
work\cite{10,11} on localization in a one-dimensional ring in the
presence of a non-hermitian field, in the context of depinning of
flux lines bound to columnar defects in a superconductor\cite{12}.

The one-dimensional disordered ring threaded by an A-B flux $\phi$
is modelled by a tight binding system composed of $N$ one-orbital
atomic sites, $n=1,2,\ldots,N$, of spacing $c_1$ forming a circular
lattice enclosing the A-B flux. The Schr\"{o}dinger equation for
this system reduces to the set of tight-binding equations\cite{13}

\begin{align}\label{eq1}
-e^{i\frac{2\pi}{N}\frac{\phi}{\phi_0}} \varphi_{n+1} -e ^{-
i\frac{2\pi}{N}\frac{\phi}{\phi_0}} \varphi_{n-1}
+\varepsilon_n\varphi_n
&=E\varphi_n\quad ,\\
n &=2,3,\ldots ,N-1 \quad ,\nonumber
\end{align}

\begin{equation}\label{eq2}
-e^{i\frac{2\pi}{N}\frac{\phi}{\phi_0}} \varphi_{2} -e
^{-i\frac{2\pi}{N}\frac{\phi}{\phi_0}} \varphi_{N}
+\varepsilon_1\varphi_1 =E\varphi_1\quad ,
\end{equation}

\begin{equation}\label{eq3}
-e^{i\frac{2\pi}{N}\frac{\phi}{\phi_0}} \varphi_{1} -e
^{-i\frac{2\pi}{N}\frac{\phi}{\phi_0}} \varphi_{N-1}
+\varepsilon_N\varphi_N =E\varphi_N\quad ,
\end{equation}
where $\varphi_n$ denotes the amplitude of an eigenstate
wavefunction at site $n$, $E$ and $\varepsilon_m$ ($m=1,2,\ldots N$)
are the corresponding eigenvalue and random site energies in units
of a fixed nearest-neighbour hopping parameter.  The flux-dependent
phase factors in (\ref{eq1}-\ref{eq3}) (where $\varphi_0=hc/e$ is
the flux quantum, with $h$ the Planck constant, $c$ the speed of
light and $-e$ the electron charge) describe exactly\cite{14} the
effect of the flux-modified boundary conditions by which the effect
of an A-B flux on the wave function of a tight-binding ring is
generally introduced\cite{1,15}.

The inverse localization length, $\xi^{-1}$, of states of energy $E$
of the ring threaded by the flux $\phi$
 may be defined by\cite{9}

\begin{align}\label{eq4}
\frac{1}{\xi} &= \lim_{n\rightarrow\infty}\frac{1}{nc_1} \langle
\ln\varphi_n
\rangle\quad , \nonumber\\
&= \lim_{n\leq N\rightarrow\infty}\frac{1}{nc_1} \sum^{n}_{p=2}
\langle \ln |R_p|\rangle\quad ,
\end{align}
where we have defined the amplitude ratios (Riccati variables)
$R_n=\varphi_n/\varphi_{n-1}, n=2,3,\ldots N$ and
$Q_1=\varphi_1/\varphi_N$ in terms of which (\ref{eq1}-\ref{eq3})
take a convenient form for recursive solution\cite{10}, starting
from an arbitrary $\varphi_1$ at the initial site $n=1$.  The
angular brackets in \eqref{eq4} denote averaging over the disorder.

Detailed analytic solutions of Eqs. (\ref{eq1}-\ref{eq3}), rewritten
in terms of the Riccati variables and for a non hermitian field $h$
substituted for $\frac{i2\pi \phi}{c_1\phi_0}$

\begin{equation}\label{eq5}
h \leftrightarrow \frac{i2\pi\phi}{c_1\phi_0}\quad ,
\end{equation}
have been discussed previously\cite{10} to second order in the
random sites energies, for arbitraty magnitude of $h$.  In Ref. 10
we performed the averaging of $\ln R_n$ expressed in terms of
zeroth-, first- and second order contributions, $R^{(0)}_n,
R^{(1)}_n$ and $R^{(2)}_n $, in $R_n$, namely

\begin{multline}\label{eq6}
\ln |R_n|=\frac{1}{2}\left(e^{-iq}R^{(1)}_n+c.c.\right)\\
+\frac{1}{2}\left[e^{-iq}\left(R^{(2)}_n-\frac{e^{-iq}}{2}
R^{(1)2}_n \right)+c.c.\right], n=2,3,\ldots N \quad ,
\end{multline}
where

\begin{equation}\label{eq7}
R^{(0)}_n=e^{iq} \quad ,
\end{equation}
is related to the energy by

\begin{equation}\label{eq8}
E=2\cos \left(q+\frac{2\pi}{N}\frac{\phi}{\phi_0}\right)\quad ,
\end{equation}
(where eigenstate energies in the absence of disorder correspond to
values $q=\frac{2\pi}{N}k, k=0,1,2\ldots$), assuming the site
energies to be identically distributed idenpendent Gaussian
variables:

\begin{equation}\label{eq9}
\langle\varepsilon_i\rangle=0,\langle\varepsilon_i\varepsilon_j\rangle=
\varepsilon^2_0 \delta_{i,j}, i=1,2,\ldots , N\quad .
\end{equation}

We refer to\cite{10} for the final explicit results for $\langle
R^{(1)2}_n \rangle$ and $\langle R^{(2)}_n \rangle$ (in the
non-hermitian field case), using the identification\eqref{eq5}.  The
inverse localization length \eqref{eq4} of the eigenstates of the
ring threaded by the flux $\phi$ obtained from Eq (21) of Ref. 10 is
given by

\begin{equation}\label{eq10}
\frac{1}{\xi}= -\frac{\varepsilon^2_0}{4}
\frac{a}{(1-a)^2}\lim_{N\rightarrow\infty}
\left(\frac{1+a^N}{1-a^N}\right)+c.c.\quad ,
\end{equation}
where

\begin{equation}\label{eq11}
a=e^{-2i\left(q+\frac{2\pi}{N}\frac{\phi}{\phi_0}\right)}\quad .
\end{equation}

The analysis of (10) in Ref. 10 for the case of an imaginary vector
potential corresponding to the real field parameter $h$ leads to a
finite so-called inverse directed localization length (IDLL) at
energy $E$ (Eq. (31) in Ref. 10).  In contrast, in the present case
we obtain from (\ref{eq8}-\ref{eq11})

\begin{align}\label{eq12}
\frac{1}{\xi}&=-i \frac{\varepsilon^2_0}{4}\;
\frac{\cot \left(qN +\frac{2\pi\phi}{\phi_0}\right)}{4-E^2}+c.c.\quad ,\nonumber\\
&=0\quad ,
\end{align}
which shows that the solutions of the Scr\"{o}dinger equations
(\ref{eq1}-\ref{eq3}) in the presence of an A-B flux are delocalized
at order $\varepsilon^2_0$ i.e. we expect that $\frac{1}{\xi}=O
(\varepsilon^4_0)$.  Since this result is obtained for any $E$ we
find, in particular, that the exact eigenstates of the ring threaded
by an A-B flux have strongly enhanced localization lengths.

This contrasts with the exact result for the inverse localization
length for weak disorder in a linear chain ($\phi=0$) given by the
celebrated Thouless formula\cite{16,17,18,19}, namely

\begin{equation}\label{eq13}
\frac{1}{\xi_1}=\frac{\varepsilon^2_0}{2}\frac{1}{4-E^2}\quad ,
\end{equation}
which is of order $\varepsilon^2_0$.  We recall that \eqref{eq13}
has also been rederived in Ref. 10, using Eq. \eqref{eq4}.  In this
case the validity of the result is supported by the general theorems
of F\"{u}rstenberg and of Oseledec\cite{9} on the properties of
infinite products of independent random 2x2 (transfer) matrices, in
terms of which the inverse localization length of an unbounded
linear chain may be defined.  The localization length in a
one-dimensional disordered ring at $\phi=0$ has sometimes been
identified with the Thouless expression \eqref{eq13} for a linear
chain (see e.g. \cite{22}).  However, no valid proof for this
identification has been provided.  As a result, there exists at
present no analytic theory of localization on a weakly disordered
one-dimensional ring at zero flux.  In contrast, the present work
based on Ref.\cite{10} does provide a theory of localization for the
case of a ring threaded by a finite A-B flux.

Our result \eqref{eq12} suggests, in particular, that persistent
current in weakly disordered metallic rings threaded by an AB-flux
is carried by states which are delocalized (i.e. free electron-like)
to order $\varepsilon^{-2}_0$.  The delocalized states (to the order
$\varepsilon^{-2}_0$) induced by the AB-flux imply the existence of
a quasi-metallic domain for ring perimeters lying between the
elastic scattering mean free path, $\ell$, and the localization
length, $\xi >>\ell$, at finite flux.  In the absence of the flux
such a domain does not exist since $\xi$ is then of the order of
$\ell$ for a one-dimensional system\cite{20}.

Finally, we observe that in our recent study of persistent current
in the tight-binding ring described by
(\ref{eq1}-\ref{eq3})\cite{14}, the effect of a weak disorder enters
generally via an overall renormalization factor of the free electron
current in the absence of disorder.  Similar conclusions have also
been reached in the study of persistent current in the more general
case of spatially correlated random potentials\cite{21}.  In this
sense our finding that the eigenstates of a weakly disordered ring
threaded by an A-B flux are delocalized to order
$\varepsilon^{-2}_0$ is consistent with our earlier results for
persistent current\cite{14,21}.


\begin{thebibliography}{unsrt}
\section{REFERENCES}
\bibitem{1} M. B\"{u}ttiker, Y. Imry, and R. Landauer, Phys. Lett. {\bf96A}, 365 (1983).
\bibitem{2}L.P. L\'{e}vy, G. Dolan, J. Dunsmuir and H. Bouchiat, Phys. Rev. Lett. {\bf64}, 2074 (1990).
\bibitem{3} V. Chandraschkar; R.A. Webb, M.J. Brady, M.B. Ketchen, W.J. Gallagher, and A. Kleinsasser, Phys. Rev. Lett. {\bf67}, 3578 (1991).
\bibitem{4} D. Mailly, C. Chapelier and A. Benoit, Phys. Rev. Lett. {\bf70}, 2020 (1993).
\bibitem{5} E.M.Q. Jariwala, P. Mohanty, M.B. Ketchen, and R.A. Webb, Phys. Rev. Lett. {\bf86}, 1594 (2001).
\bibitem{6} R. Deblock, R. Bel, B. Reulet, H. Bouchiat, and D. Mailly, Phys. Rev. Lett. {\bf89}, 206803 (2002).
\bibitem{7} Y. Imry, Introduction to Mesoscopic Physics (Oxford University Press, Oxford, 1997).
\bibitem{8} K. Ishii, Prog. Theor. Phys., Suppl. {\bf53}, 77 (1973).
\bibitem{9} A. Crisanti, G. Paladin and A. Vulpiani, Products of Random Matrices in Statistical Physics (Springer, Berlin, 1993).
\bibitem{10} J. Heinrichs, Phys. Stat. Sol.(b) {\bf231},19 (2002).
\bibitem{11} J. Heinrichs, Pramana-J. Phys. {\bf58}, 153-171 (2002) (Proceedings of Int. Discussion Meeting on Mesoscopic and Disordered Systems, Bangalore, 18-20 December 2000, Ed. A.M. Jayannavar, H.R. Krishnamurthy and A.K. Raychariduri).
\bibitem{12} N. Hatano and D.R. Nelson, Phys. Rev. Lett. {\bf77}, 570 (1996).
\bibitem{13} H.F. Cheung, Y. Gefen, E.K. Riedel and W.H. Shih, Phys. Rev. B{\bf37}, 6050 (1988).
\bibitem{14} J. Heinrichs, Int. J. Mod. Phys. B{\bf16}, 593 (2002).
\bibitem{15} N. Byers and C.N. Yang, Phys. Rev. Lett. {\bf7}, 46 (1961).
\bibitem{16} D.J. Thouless, in Ill-Condensed Matter, Edts R. Balian, R. Maynard and G. Toulouse, North Holland, Amsterdam, 1979.
\bibitem{17} J.M. Luck, Les Syst\`{e}mes D\'{e}sordonn\'{e}s Unidimensionnels, 1992 (Collection Al\'{e}a-Saclay)(Saclay:Commissariat de l'Energie Atomique).
\bibitem{18} B. Kramer and A. MacKinon, Rep. Prog. Phys. {\bf56}, 1469 (1993).
\bibitem{19} J. Pendry, Adv. Phys. {\bf43}, 461 (1994).
\bibitem{22} P.W. Brouwer, P.G. Silvestrov, and W.C.J. Beenakker, Phys. Rev. B {\bf 56}, R4333 (1997).
\bibitem{20} D.J. Thouless, J. Phys. C{\bf6}, 649 (1973).
\bibitem{21}J. Heinrichs, J. Phys.: Condensed Matter {\bf20}, 245232 (2008).
\end{thebibliography}
\end{document}